\documentclass[12pt]{iopart}
\usepackage[english]{babel}
\usepackage[dvips]{graphicx}
\usepackage{amsmath}
\usepackage{amsfonts}
\usepackage{amssymb}
\usepackage{verbatim}

\begin{document}
\title[]{Suppression of stochastic fluctuations of suspended 
nanowires by temperature-induced single-electron tunnelling}

\author{F Santandrea$^1$, L Y Gorelik$^2$, R I Shekhter$^1$ and 
M. Jonson$^{1,3,4}$}
\address{$^1$ Department of Physics, University of Gothenburg, SE - 412 96
G{\"o}teborg, Sweden}
\address{$^2$ Department of Applied Physics, Chalmers University of Technology,
SE - 412 96 G{\"o}teborg, Sweden}
\address{$^3$ SUPA, Department of Physics, Heriot-Watt University,
Edinburgh EH14 4AS, Scotland, UK}
\address{$^4$Division of Quantum Phases and Devices, School of Physics,
Konkuk University, Seoul 143-107, Korea}
\ead{gorelik@chalmers.se}

\begin{abstract}
We investigate theoretically the electromechanical properties of freely 
suspended nanowires that are in tunnelling contact with the tip of a scanning 
tunnelling microscope (STM) and two supporting metallic leads. The aim of our 
analysis is to characterize the fluctuations of the dynamical variables of the
nanowire when a temperature drop is mantained between the STM tip and the 
leads, which are all assumed to be electrically grounded. By solving a quantum 
master equation that describes the coupled dynamics of electronic and 
mechanical degrees of freedom we find that the stationary state of the 
mechanical oscillator has a Gaussian character, but that the amplitude of 
its root-mean square center-of-mass fluctuations is smaller than would be 
expected if the system were coupled only to the leads at thermal equilibrium. 
%This effect can be interpreted as an effective cooling of the mechanical 
%degrees of freedom induced by the nonequilibrium electronic environment.
\end{abstract}
\submitto{\NJP}
\maketitle

%\noindent
\section{Introduction}
The possibility to detect and control the motion of nanometer-sized-mechanical 
resonators by coupling them to mesoscopic electronic devices has generated a 
considerable research effort in recent years \cite{Blencowe2005}. 
In particular, the possibility to use such nanoelectromechanical systems (NEMS)
as ultrasensitive sensors of, for example, displacement and mass,
have been demonstrated in a number of works \cite{Teufel2009,Naik2009}.
Independently of the specific type of electronic device considered in the 
different schemes, a common feature that has emerged from these studies is that 
the electronic subsystem must be out of thermodynamic equilibrium in 
order to function as an ultrasensitive measurement tool in combination with the
mechanical subsystem. This observation naturally raises the question of
how the dynamics of the mechanical subsystem is affected by the 
\textit{nonequilibrium environment} created by the mesoscopic electronic 
device to which it is coupled. 

It is known from statistical mechanics that the displacement and momentum 
fluctuations of a quantum harmonic oscillator (the basic model for any movable
structure that could be included in a NEMS), which is coupled to a thermal bath
in equilibrium at temperature $T$ are described by Gaussian distribution
functions, whose widths are proportional to $\coth^{1/2}(\hbar \omega/2k_BT)$ 
where $\omega$ is the frequency of the oscillator. From this formula it follows
that the fluctuations have a thermodynamic origin in the high-temperature 
limit, where they are fully defined by the temperature of the thermal bath. On 
the other hand, at low temperatures $k_BT \ll \hbar \omega$ the fluctuations are
completely quantum mechanical in nature. 

What kind of changes from this picture could be expected if the oscillator is 
coupled to a \textit{nonequilibrium} environment? In spite of the difficulties 
related to the definition of temperature for systems that are out of 
thermodynamic equilibrium, several theoretical works show that  
nonequilibrium fluctuations in the properties of nanomechanical oscillators 
coupled to mesoscopic electronic systems (such as a single-electron
transistor or a superconducting Cooper pair box) are, to a good approximation, 
still described by Gaussian distribution functions 
\cite{Armour2002,Armour2004}.

A remarkable difference between a passive (thermodynamic) environment and 
an active (nonequilibrium) one is that in the latter case the amplitude of 
the fluctuations can be controlled through parameters that characterize
the state of the electronic subsystem. This external control introduced by 
the coupling between mechanical and electronic degrees of freedom opens the 
way for the possibility to reach the quantum limit of fluctuations even when 
the temperature is high on the scale defined by the quantum of mechanical 
energy, i.e. $\hbar \omega /k_B$. 

In the last few years considerable efforts have been made in order to, 
develop efficient procedures to effectively ``cool'' down the motion of 
nanomechanical resonators below the threshold defined by thermal
fluctuations. Most of the proposed schemes strive to reproduce the effects of 
laser cooling of atoms and molecules by purely electronic means 
\cite{Zippilli2009,Pistolesi2009,Ouyang2009,Sonne2010}. 
The general strategy underlying these approaches is based on the coherent 
control of resonant, energy-conserving transitions between \textit{discrete} 
electronic levels. 

Recently we have proposed a fundamentally new scheme to cooling the 
vibrations of a suspended-nanowire based mechanical oscillator 
\cite{Santandrea2011}. In contrast to the aforementioned cooling schemes, 
our proposal has the advantage that it does not rely on the energy 
conservation constraint. In particular, we considered a suspended carbon 
nanotube in tunnelling contact with the voltage-biased tip of a scanning 
tunnelling microscope (STM) and the metallic leads at which its ends are 
clamped. Our analysis showed that the average number of quantized vibrational 
excitations, i.e. vibrons, (which is proportional to the root-mean square 
fluctuations of the center-of-mass position) can be reduced by varying the 
bias voltage within a range of values for which the probability for absorbing 
vibrons during inelastic electron tunnelling processes is significantly 
enhanced over the probability for vibron emission. 

In this paper we demonstrate that the cooling mechanism suggested in Ref. 
10 can work by exploiting the temperature gradient, rather 
than the potential drop across the system. We show that the amplitude of the 
root-mean square fluctuations of the nanotube center-of-mass position is 
smaller than what would be in presence of only an equilibrium thermal bath. 
This partial suppression of the stochastic fluctuations of the nanotube 
displacements can be interpreted as an effective cooling of the mechanical 
degrees of freedom of the system. Moreover, we have found that this effective 
cooling phenomenon involves simultaneously several low-frequency vibrational 
modes and not only the fundamental one. 

\section{Model Hamiltonian}
\begin{figure}
\center
\includegraphics[width=0.5\textwidth]{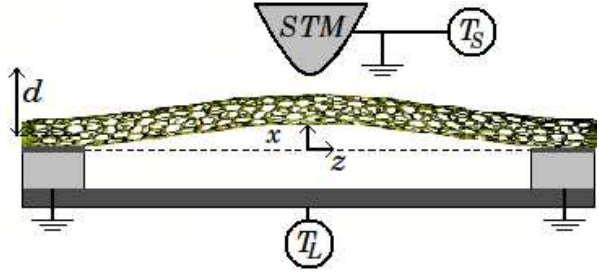}
\caption{Sketch of the model system considered. A carbon nanotube is suspended 
over a trench between metallic leads, while an STM tip is placed a distance $d$
above the nanotube. Both the STM and the leads are grounded, while their 
temperatures $T_S$ and $T_L$ are different and held constant ($T_S>T_L$ is
assumed).} 
\label{fig:system}
\end{figure}

To be specific, we consider the system sketched in Fig.~\ref{fig:system}, where 
a carbon nanotube is suspended over a trench between two metallic leads. Its 
ends are both clamped, while a third electrode in the form of the tip a 
scanning tunnelling microscope (STM) is positioned above the nanotube. 
The suspended carbon nanotube can be considered as a quantum dot that is
coupled to the surrounding electrodes through tunnel junctions. 
Low-temperature tunnelling spectroscopy studies on freely hanging carbon 
nanotubes have shown that inelastic electron tunnelling can create a non-thermal
equilibrium population of the vibronic states of the nanotube \cite{LeRoy2004}.
%------------------------------------
%-----------------------------------

In order to analyze the dynamics of the nanotube deflections and the behavior
of the electronic subsystem in the quantum regime we introduce a model 
Hamiltonian, $H= H_e+H_m+H_T+H_C$, where the different contributions are 
given by:
\begin{subequations}
\begin{align} 
\label{def:H_e}
H_e & =\sum_{q,\alpha} E_{q,\alpha} a_{q,\alpha}^{\dagger} a_{q,\alpha}+E_0c^\dagger c \,,\\
\label{def:H_m}
H_m & = \int_{-L/2}^{L/2}dz\left\{\frac{\hat{\pi}^2(z)}{2\rho}+
\frac{\kappa [\hat{u}''(z)]^2}{2}\right\}\,,\\
\label{def:H_T}
H_T & =\sum_q \left\{ t_S[\hat{u}(z_0)]c^\dag a_{q,S} + t_L
a_{q,L}^\dagger c \right\}+\rm{H.c.},\\
\label{def:H_C}
H_C & =-\Im \hat{u}(z_0)c^\dagger c.
\end{align}
\end{subequations}
\noindent In Eqs. (\ref{def:H_e}), (\ref{def:H_T}) and (\ref{def:H_C}), 
$a_{q,\alpha}^{(\dagger)}$ and $c^{(\dagger)}$ are annihilation (creation) operators 
for electrons in the STM tip ($\alpha=S$), in the leads ($\alpha=L$) and in 
the nanotube, respectively. The term $H_e$ in Eq. (\ref{def:H_e}) describes the 
electronic states in the STM tip, the leads (which are treated as reservoirs of 
non-interacting quasiparticles) and in the nanotube. We assume that, in the 
range of temperatures that are relevant for our considerations, only one 
quantized electronic level in the nanotube is involved in the exchange of 
charge with the reservoirs and we denote its energy as $E_0$. Such a condition 
is satisfied if the temperatures of the reservoirs are significantly lower 
than the characteristic difference in energy between the quantized electronic 
levels of the nanotube, which can be estimated as 
$\Delta \simeq \hbar v_F/2\ell \approx$ 1.7 meV/$\mu$m, where $v_F$ is Fermi 
velocity and $\ell$ is the length of the nanotube.

The term $H_m$ in Eq. (\ref{def:H_m}) describes the mechanical degrees of 
freedom of the nanotube. The quantum field $\hat{u}(z)$ gives the nanotube 
deflection from the straight configuration at point $z$ (that is, the nanotube 
axis, see Fig. \ref{fig:system}), while $\hat{\pi}(z)$ is the momentum linear 
density and the symbol ${}'$ denotes derivation with respect to the coordinate 
$z$. The displacement and momentum density fields are canonically conjugated 
dynamical variables, that is they obey the commutation relation: 
$[\hat{u}(z_i),\hat{\pi}(z_j)]=i\hbar\delta(z_i-z_j)$. 

The parameter $\rho$ represents the linear mass density of the nanotube, 
$\kappa$ its bending rigidity, and $L$ is the length of the suspended part 
(notice that $L \neq \ell$, where the latter is the total length of the 
nanotube). The clamping of both the ends of the nanotube to the leads can be 
expressed through the boundary conditions 
$\hat{u}(\pm L/2)=\hat{u}'(\pm L/2)=0$.

The tunnelling of electrons through the STM tip-nanotube and the
nanotube-leads junctions is described by the Hamiltonian operator $H_T$, 
presented in Eq. (\ref{def:H_T}). We denote by $z_0$ the point along the
nanotube axis above which the STM is positioned. Both the 
tunnelling amplitudes $t_S[\hat{u}(z_0)]$ and $t_L$ are assumed to be 
independent of the electronic energy, whereas only $t_S[\hat{u}(z_0)]$ is a 
function of the nanotube deflection, as a consequence of its dependence on the 
overlap of the electronic wavefunctions in the STM tip and the nanotube. We 
model this deflection dependence of the probability amplitude of tunnelling
between the STM tip and the nanotube as 
$t_S[\hat{u}(z_0)] \equiv t_S\exp[u(z_0)/\lambda]$, where $\lambda$ is the 
characteristic tunnelling length of the junction ($\lambda \sim 10^{-10}$ m). 

The effect that the nanotube displacement has on the width of the tunnel 
barrier provides a mechanism of coupling the electronic and the 
mechanical degrees of freedom of the system. In the following, we will refer 
to that as \textit{tunnelling} electromechanical coupling.

The last term in the Hamiltonian, $H_C$, shown in Eq. (\ref{def:H_C}), describes
the electrostatic interaction in the system. Since both the STM tip and the 
substrate are grounded, an electron occupying 
the state inside nanotube induces a polarization charge of opposite sign in 
the STM tip and hence generates an electrostatic force $\Im$ acting on the 
nanotube. This force can be thought as applied to the point $z_0$ and directed 
towards STM tip (this ``strongly localized'' form of the electrostatic force 
is a valid approximation if the effective radius of the STM tip is negligible 
with respect to the length of the nanotube). 

The electrostatic interaction described by the operator $H_C$ provides another 
mechanism that couples the dynamics of the mechanical and electronic degrees 
of freedom, which we will refer to hereafter as the \textit{polaronic} 
electromechanical coupling, because of the formal analogy with the interaction 
term in Hamiltonian of the polaron problem. Under the assumptions of uniform 
charge distribution inside the charged nanotube, if the equilibrium distance 
between the nanotube and the STM tip $d$ is much less than the effective radius
of the tip, $R$, then $\Im$ can be approximated by the following expression:
\begin{equation}\label{eq:el_force-zerobias}
\Im = \frac{\beta}{\varepsilon_0} \left( \frac{R}{\ell^2} \right) \frac{e^2}{d},
\end{equation}
where $\beta$ is a numerical factor of the order of one and $\varepsilon_0$
the vacuum permittivity.
From the considerations presented above, it follows that the 
electronic and mechanical subsystems interact through two independent 
coupling mechanisms, that is the tunnelling and the polaronic. 
The former one results in the change of the nanotube momentum on the value 
$\hbar/\lambda$ when one electron tunnels from the STM to the nanotube or in 
the opposite direction. The second one accounts the difference between the 
equilibrium configurations of the charged and neutral nanotube. Working 
incoherently these two mechanisms of electromechanical coupling give 
rise to stochastic fluctuations of the nanotube center-of-mass position.

However, as it was shown in Ref. 10, the quantum interplay between them may 
significantly modify their mutual performance. In order to analyze the 
consequences of the combination of the two coupling mechanisms, it is 
convenient to introduce the eigenmode representation for the nanotube 
displacement and momentum density fields, that is defined by the operators
\begin{subequations}
\begin{align} 
\label{def:X_n}
\hat{X}_n & =\frac{1}{\sqrt{L}}\int_{-L/2}^{L/2} dz \varphi_n(z)\hat{u}(z) \\
\label{def:P_n}
\hat{P}_n & =\sqrt{L}\int_{-L/2}^{L/2} dz \varphi_n(z)\hat{\pi}(z).
\end{align}
\end{subequations}
\noindent The operators presented in Eqs. (\ref{def:X_n}) and (\ref{def:P_n})
satisfy canonical commutation relations, that is 
$[\hat{X}_n,\hat{P}_l]=i\hbar\delta_{n,l}$ and the complete set of orthonormal 
functions $\varphi_n(z)$ is given by the eigenfunctions of the operator 
$d^{4}/dz^{4}$ with the boundary conditions 
$\varphi_n(\pm L/2)=\varphi_n'(\pm L/2)$. In this representation the terms 
$H_m, H_T$ and $H_C$ in the Hamiltonian assumes the form:
\begin{align}
H_m+H_C & = \sum_n\left(\frac{1}{2M}\hat{P}_n^2 + \frac{\omega^2_nM}{2}
\hat{X}_n^2\right)-\Im c^\dagger c\sum_n\varphi_n(z_0)\hat{X}_{n}\,,\\
H_T & =\sum_q \left[ t_Se^{\sum_n\varphi_n(z_0)\hat{X}_n/\lambda} c^\dag a_{q,S} + t_L
a_{q,L}^\dagger c \right] + \rm{H.c.},
\end{align}
\noindent where $M=\rho L$ is the mass of the suspended part and $\omega_n$ 
is the eigenfrequency of the $n$-th bending mode. The polaronic term can be
removed from the Hamiltonian by a suitable unitary transformation, 
$H\rightarrow \tilde{H}=UHU^{\dagger}$, where 
$U \equiv \exp[i\hbar^{-1}\Im c^\dagger c\sum_n \varphi_n(z_0)\hat{P}_n/
2M\omega^2_n]$. However, as additional consequence of this transformation, 
the tunnelling amplitudes turn out to be dependent on the momentum 
operators. The transformed tunnelling Hamiltonian reads
\begin{equation} \label{eq:transf_Htunn2}
\sum_q \left[ t_S 
e^{\sum_n\varphi_n(z_0)\left(\frac{\hat{X}_n}{\lambda}+i\Im\frac{\hat{P}_n}{2M\hbar\omega^2_n}
\right)} c^\dag a_{q,S} + 
t_{L} e^{i\sum_n\varphi_n(z_0)\left(\Im\frac{\hat{P}_n}{2M\hbar\omega^2_n}\right)}c^\dag a_{q,S}
\right] + \rm{H.c.},\\
\end{equation}
\noindent The physical analysis of Eq. (\ref{eq:transf_Htunn2}) becomes
more transparent after introducing the creation (annihilation) operators 
$b^{\dagger}_n (b_n)$ for the elementary mechanical excitations 
(vibrons) of the  $n$-th bending mode: 
$\hat{X}_n=(b^{\dagger}_n+b_n)\chi_n/\sqrt{2}$, 
$\hat{P}_n=i\hbar(\sqrt{2}\chi_n)^{-1}(b^{\dagger}_n-b_n)$, where 
$\chi_n/\sqrt{2}=\sqrt{\hbar/2M\omega_n}$ is the position uncertainty in the 
vibrational ground state of the oscillator associated to the $n$-th mode. In 
this representation, the part of the Hamiltonian which describes  the electron 
tunnelling processes between the STM tip and nanotube assume the form: 
\begin{equation} \label{eq:transf_Htunn3}
\sum_q  t_S \left\{ e^{\sum_n(A_n^+b^{\dagger}_n+A_n^-b_n)}c^\dag a_{q,S} + 
e^{\sum_n(A_n^-b^{\dagger}_n+A_n^+b_n)}a_{q,S}^{\dag}c\right\}, 
\end{equation}
\noindent where the parameters $A_n^{\pm}$, which characterize the rates of 
the inelastic electronic transitions with absorption and emission of the 
vibronic quanta, are given by:
\begin{equation} \label{def:inelastic_rates} 
A_n^{\pm}=\frac{\varphi_n(z_0)\chi_n}{\sqrt{2}}
\left(\frac{1}{\lambda} \mp \frac{\Im}{\hbar \omega_n}\right).
\end{equation} 
\noindent From the expressions of the parameters $A_n^{\pm}$ shown in 
Eq. (\ref{def:inelastic_rates}), it follows that the vibron emission processes 
are suppressed with respect to the absorption ones when electrons tunnel from 
the STM tip to the nanotube, whereas the absorption is suppressed and the 
emission promoted during the transitions in the opposite direction (which are 
described by the rightmost term of the transformed tunnelling Hamiltonian 
shown in Eq. (\ref{eq:transf_Htunn3})). 

Moreover, one can achieve complete suppression of the vibron emission 
(absorption) for the given mode by varying the value of the electrostatic force 
$\Im$, which, according to Eq. (\ref{eq:el_force-zerobias}), is controlled by 
the equilibrium distance between the STM tip and the nanotube, $d$. If after 
tunnelling from the STM tip, the electrons tunnel immediately off to the leads, 
the rate at which vibron emission processes occur will be substantially 
reduced and therefore, in the stationary regime, one can expect that the number
of vibrons will be close to zero. 

However, in order to drive a certain vibrational mode to its quantum ground 
state (which corresponds to an average number of vibrons much smaller than 1),
its frequency must satisfy the condition $\omega_n=\sqrt{2}\lambda\Im/\hbar$.
In the stationary regime, the root-mean-squared deviation of the center-of-mass 
position (which expresses the fluctuations of the mechanical state of the 
nanotube around its equilibrium configuration) is given by the square root of 
the following expression:
\begin{equation} \label{eq:CM_fluctuations}   
\langle (\hat{u}(0)-\langle \hat{u}^2(0)\rangle)^2 \rangle=
\chi_0^2\sum_n\left(\frac{\omega_0}{\omega_n}\right) \varphi_n^2(0)
\left(\langle b^{\dagger}_nb_n\rangle +\frac{1}{2}\right).
\end{equation}
\noindent Taking into account that for the doubly clamped nanotube
$\omega_0/\omega_n\simeq (n+1)^{-2}$ and that $\varphi_{2n+1}(0)=0$ 
Eq. (\ref{eq:CM_fluctuations}) indicates that the of center-of-mass 
fluctuations are mainly defined by the average number of the vibrons in the 
fundamental mode. Therefore, in order to suppress such fluctuations, this 
number shoud have the minimum possible value. On the basis of these 
considerations, in the rest of this paper we restrict our attention to 
the fluctuations of the fundamental bending mode, which will be considered 
as a quantum harmonic oscillator.

\section{Quantum master equation}

In order to perform a quantitative analysis of the phenomena discussed
above, we start from the Lioville-von Neumann equation for the density matrix 
operator, which represents the state of the whole system
\begin{equation} \label{eq:LvN_total}
i\hbar \frac{d\rho}{dt} = [H,\rho(t)].
\end{equation}
In Born approximation with respect to the tunnelling amplitudes $t_S$, $t_L$, 
Eq. (\ref{eq:LvN_total}) can be recast in the following integral form:
\begin{equation}\label{eq:LvN_int-diff}
\frac{d\rho}{dt} = \frac{1}{i\hbar}[H_0,\rho]
-\frac{1}{\hbar^2}\int_{-\infty}^tdt'[H_T(t),[H_T(t'),\rho(t')]].
\end{equation}
\noindent where $H_0 \equiv H_e+H_m+H_C$, and 
$\hat{\mathcal{A}}(t) = e^{iH_0t/\hbar}\hat{\mathcal{A}}e^{-iH_0t/\hbar}$. Taking 
into account that the coupling between the nanotube and the electronic 
reservoirs (that is, the STM tip and the leads) is weak enough so that any 
back-action of the nanotube on their physical states is negligible, we can
use the Ansatz: $\rho(t) \approx \sigma(t) \otimes \rho_S \otimes \rho_L$.
The operator $\sigma(t)$ is the \textit{reduced} density
matrix operator, which is defined as $\sigma(t) \equiv \rm{Tr}_{S+L}[\rho(t)]$
and represents the electronic and mechanical state of the oscillator. The 
density matrices $\rho_\alpha$ describe the STM tip ($\alpha=S$) and the leads 
($\alpha=L$) as electronic reservoirs at thermal equilibrium with temperatures 
$T_S$ and $T_L$, which means that:
\begin{displaymath}
\mathrm{Tr} (a^{\dag}_{q,\alpha}a_{q,\alpha}\rho_{\alpha}) = 
\left(1+exp\{(E_{q,\alpha)}-\mu_\alpha)/k_BT_{\alpha}\}\right)^{-1}
\equiv f_{\alpha}(E_{q,\alpha}-\mu)
\end{displaymath}
\noindent where the $\mu$ is the chemical potential of the STM tip 
and the leads, which is supposed to be the same.

In order to describe the nanotube dynamics it is convenient to project the 
reduced density matrix onto the subspaces corresponding to the electronic
level in the nanotube being occupied or unoccupied. That amounts to multiply
the operators $c^\dagger c$ and $cc^\dagger$ to Eq. (\ref{eq:LvN_int-diff}) and  
trace over the electronic degrees of freedom of the nanotube. After this 
procedure, we obtain two coupled equations for the operators 
$\sigma_0 \equiv \rm{Tr}_e(\sigma cc^\dagger)$ and 
$\sigma_1 \equiv \rm{Tr}_e(\sigma c^\dagger c)$ which, in the high-temperature 
limit $\hbar \omega /k_BT_L \ll 1$ turn out to be local in time 
\cite{Zazunov2006}. For small displacements of the nanotube around the 
equilibrium configuration, the tunnelling amplitude $t_S[\hat{X}]$ can be
linearized, so that the equations of motion for $\sigma_{0,1}$ have the form:
\begin{subequations}
\begin{align} 
\label{eq:rho1_Markov}
\frac{d\sigma_1}{dt} & = -\frac{i}{\hbar}[H_m,\sigma_1] +
\frac{i\Im}{\hbar}[\hat{X},\sigma_1]-\Gamma_L^-\sigma_1 
- \Gamma_S^-\left(\sigma_1+\frac{1}{\lambda}\{ \sigma_1,\hat{X} \}+
\frac{1}{\lambda^2}\{ \sigma_1,\hat{X}^2 \}\right)+ \nonumber \\
{} & + \Gamma_S^+\left(\sigma_0+\frac{1}{\lambda}\{ \sigma_0,\hat{X} \}+
\frac{1}{2\lambda^2}\{ \sigma_0,\hat{X}^2 \} +
\frac{1}{\lambda^2} \hat{X}\sigma_0 \hat{X}\right) + \Gamma_L^+\sigma_0
+\mathcal{L}_\gamma\sigma_1\\
%--------------------------------------
\label{eq:rho0_Markov}
\frac{d\sigma_0}{dt} & = -\frac{i}{\hbar}[H_m,\sigma_0] +
\Gamma_L^-\sigma_1 +\Gamma_S^-\left(\sigma_1+
\frac{1}{\lambda}\{ \sigma_1,\hat{X} \}+
\frac{1}{2\lambda^2}\{ \sigma_1,\hat{X}^2 \} +
\frac{1}{\lambda^2} \hat{X}\sigma_1 \hat{X}\right)- \nonumber \\
{} & -\Gamma_S^+\left(\sigma_0+\frac{1}{\lambda}\{ \sigma_0,\hat{X} \}+
\frac{1}{\lambda^2}\{ \sigma_0,\hat{X}^2 \}\right) - \Gamma_L^+\sigma_0
+\mathcal{L}_\gamma\sigma_0,
\end{align}
\end{subequations}
\noindent where $\Gamma_\alpha^+ \equiv \Gamma_\alpha f_\alpha(E_0)$,
$\Gamma_\alpha^- \equiv \Gamma_\alpha [1-f_\alpha(E_0)]$ 
and $\Gamma_\alpha \equiv 2\pi \hbar^{-1}|t_\alpha|^2\nu_\alpha$, $\nu_\alpha$
being the density of states at the Fermi energy in the STM tip 
($\alpha = S$) and in the leads ($\alpha = L$).

The operator $\mathcal{L}_\gamma$ in Eqs. (\ref{eq:rho1_Markov}) and
(\ref{eq:rho0_Markov}) models the relaxation of the oscillator towards thermal 
equilibrium with the phononic bath in the leads, a process characterized by the 
rate $\gamma \equiv \omega_0 /Q$, where $Q$ is the quality factor. On the basis 
of general considerations regarding quantum dissipative systems 
\cite{Breuer2002}, $\mathcal{L}_\gamma$ can be explicitly written as:
\begin{equation} \label{def:dissipative_op}
\mathcal{L}_\gamma[\sigma] \equiv 
-\frac{i\gamma}{2\hbar}[\hat{X},\{\hat{P},\sigma \}]-
\frac{\gamma}{2\chi_0^2} \coth ( \hbar \omega_0/2k_BT_L)
[\hat{X},[\hat{X},\sigma]].
\end{equation}
The mechanical state of the suspended nanotube can be characterized through the
operator $\sigma_+\equiv \sigma_0+\sigma_1$, whose evolution in time is fully 
determined by Eq. (\ref{eq:rho1_Markov}) and (\ref{eq:rho0_Markov}), once that
the operator $\sigma_- \equiv \sigma_0-\sigma_1$ is introduced. Furthermore, 
in order to describe the stationary state of the oscillator, it is convenient 
to introduce the dimensionless operators $\hat{x}=\hat{X} / \chi_0$,  
$\hat{p}=\chi_0 \hat{P}/\hbar$ and express the operators $\sigma_\pm$ in the 
``Wigner function representation'' \cite{Hillery1984}, which is defined as:
\begin{equation} \label{eq:Wigner_f}
W_\pm(x,p,t) = \int_{-\infty}^{+\infty}\frac{d\xi}{\pi}
e^{-i2p\xi} \langle x-\xi|\sigma_\pm(t)|x+\xi \rangle
\end{equation}
From Eqs. (\ref{eq:rho1_Markov}) and (\ref{eq:rho0_Markov}), it follows that 
the Wigner functions corresponding to the stationary solutions of the quantum 
master equations for the operators  $\sigma_+$, $\sigma_-$ satisfy the 
equations:
\begin{subequations}
\begin{align} 
\label{eq:W+}
(p\partial_x-x\partial_p)\overline{W}_+ & = 
\frac{\varepsilon_p}{2} \frac{\Delta \Gamma_\Sigma}{2\Gamma_\Sigma}
\partial_p \overline{W}_+ + \frac{\varepsilon_p}{2}\partial_p\overline{W}_- +
\varepsilon_p^2\frac{\Gamma_S}{2}\partial_p^2\overline{W}_+ +
\varepsilon_t^2\frac{\Delta \Gamma_s}{4} \partial_p^2\overline{W}_-+\nonumber \\ 
{} & + \frac{1}{Q}\partial_p(p\overline{W}_+) +
\frac{\coth(\hbar \omega_0/2k_BT_L)}{2Q}\partial_p^2\overline{W}_+ 
+\mathcal{O}(Q^{-2},\varepsilon_t^4)\overline{W}_-\\
%-----------------------------------
\label{eq:W-}
(p\partial_x-x\partial_p)\overline{W}_- & =
\varepsilon_p \frac{\Delta \Gamma_\Sigma}{2\Gamma_\Sigma}\partial_p\overline{W}_-
 + \frac{\varepsilon_p}{2} \partial_p \overline{W}_+-
\left(\Delta \Gamma_\Sigma +2\varepsilon_t \Delta \Gamma_S x \right)
\overline{W}_+ - \nonumber \\
{} & -\left(\Gamma_\Sigma +2\varepsilon_t \Gamma_S x \right)\overline{W}_-
+\mathcal{O}(Q^{-1},\varepsilon_t^2)\overline{W}_-
\end{align}
\end{subequations}
\noindent where $\varepsilon_t=\varphi_0(z_0)\chi_0/\lambda$, 
$\varepsilon_p=\varphi_0(z_0)\Im/\chi_0 M\omega_0^2$, while the parameters related
to the tunnelling processes have been rescaled in units of $\omega_0$, that is
$\Delta \Gamma_\alpha \equiv (\Gamma_\alpha^+-\Gamma_\alpha^{-})/
\omega_0$, where $\alpha=S,L$, 
$\Delta \Gamma_\Sigma=(\Delta \Gamma_S+\Delta \Gamma_L)/\omega_0$ and
$\Gamma_\Sigma=(\Gamma_S+\Gamma_L)/\omega_0$.

Eqs. (\ref{eq:W+}) and (\ref{eq:W-}) can be solved by means of 
a perturbative expansion in the small coupling constants 
$\varepsilon_p \sim \varepsilon_t \ll 1$ and inverse quality factor 
$Q^{-1} \simeq \varepsilon_p^2$, $\varepsilon_t^2$. At the zero-th order 
in the small parameters, the solution of Eqs. (\ref{eq:W+}) and (\ref{eq:W+}), 
has the form $\overline{W}_+^{(0)} = w(A)$, 
$\overline{W}_-^{(0)} = -(\Delta \Gamma_\Sigma / \Gamma_\Sigma) w(A)$,
where $A \equiv \sqrt{x^2+p^2}$ and $w$ is an arbitrary function.

The necessary and sufficient condition for $w(A)$ to be a good zero-order 
approximation of $\overline{W}_+$ is that any deviation from $w(A)$
is at most of order ($\varepsilon_t^2$, $\varepsilon_p^2$, $Q^{-1}$)
in comparison to $w(A)$. Then, by replacing the expressions of 
$\overline{W}_\pm$ up to the second-order corrections in the small parameter 
into Eqs. (\ref{eq:W+}) and (\ref{eq:W-}), and neglecting all 
the contributions except the zero-th order ones, we obtain an equation that 
determines the form of $w(A)$.   

The first order corrections to the function $\overline{W}_+^{(0)}=w(A)$ 
vanish after performing the transformation $x\rightarrow x - \bar{x}$ 
(where $\bar{x}=(\varepsilon_p/2)(1-\Delta \Gamma_\Sigma / \Gamma_\Sigma)$), which
can be considered as a shift of the reference frame that accounts for the 
nanotube deformation induced by the average electrostatic force.

\section{Results}
Following the perturbative procedure described above, it turns out that, for 
the function $w(A)$ to be an appropriate approximation for 
$\overline{W}_+$, it must satisfy the following first-order linear differential
equation:
\begin{align} \label{eq:determine_w(A)}
{} & \left( \frac{\varepsilon_p^2}{4}
\frac{\Gamma_\Sigma(1-(\Delta \Gamma_\Sigma / \Gamma_\Sigma )^2)}{1+\Gamma_\Sigma^2}
+ \frac{\varepsilon_p^2}{4}
\frac{\Gamma_S \Gamma_\Sigma-\Delta \Gamma_S \Delta \Gamma_\Sigma}{\Gamma_\Sigma}+
\frac{1}{2Q}\coth \left( \frac{\hbar \omega_0}{2k_B T_L}\right) \right)
\partial_A w = \nonumber \\
{} & -2\left(
\frac{\varepsilon_p \varepsilon_t}{2}
\frac{\Gamma_L \Delta \Gamma_S - \Gamma_S \Delta \Gamma_L}
{\Gamma_\Sigma(1+\Gamma_\Sigma^2)} +
\frac{1}{Q} \right) Aw
\end{align}
\noindent Now we focus our attention on the regime in which the transport of
charge from the STM tip to the leads is activated by the gradient of
temperature between the tip and the leads. In order to have an 
appreciable rate of tunnelling between the STM tip and the nanotube,
the STM tip temperature must satisfy $k_BT_S \sim E_0$, so that the
electronic states at the energy of the electronic level of the 
nanotube have a good chance to be populated, that is $f_S(E_0) \simeq 1/2$.
At the same time, the temperature in the leads should be much lower
than $T_S$, so that the electronic states in the leads at the energy
corresponding to the electronic level in the nanotube have a good
chance to be empty, which means $f_L(E_0) \simeq 0$. Furthermore, 
the temperature of the leads should be high on the scale defined by 
the vibrational quantum energy, that is $k_BT_L \gg \hbar \omega_0$.

For what concerns the best cooling performance, we already observed that 
it is expected to be achieved if the electrons, after having tunneled to 
the nanotube from the STM tip, tunnel quickly to the leads rather than 
being transferred back to the STM tip. That corresponds to the situation 
in which $\Gamma_S / \Gamma_L \ll 1$. Under these conditions, the 
quasi-distribution function $w(A)$ has the following Gaussian form:
\begin{subequations}
\begin{align} \label{eq:w(A)}
w(A) & = \frac{1}{\pi \theta^2}e^{-\frac{A^2}{\theta^2}}, \\
%---------------------------
\label{def:theta}
\theta^{-2} & = \frac{\frac{\varepsilon_p \varepsilon_t}{2(1+\Gamma_\Sigma^2)}+
\frac{\Gamma_\Sigma}{\Gamma_S\Gamma_L}\frac{1}{Q}}
{\frac{\varepsilon_p^2}{2(1+\Gamma_\Sigma^2)} +
\frac{\varepsilon_t^2}{4}+
\coth(\hbar \omega_0/2k_BT_L)\frac{\Gamma_\Sigma}{\Gamma_S\Gamma_L}\frac{1}{Q}}.
\end{align}
\end{subequations}
\noindent We remark that, by virtue of the Gaussian form of the stationary
state, the root-mean-square fluctuations of the nanotube center-of-mass 
position are just proportional to the width of the quasi-distribution function,
i. e. $\langle X^2 \rangle^{1/2} = \theta / \sqrt{2}$.

For an oscillator coupled to an equilibrium environment, in the 
high-temperature limit, $\theta^2$ becomes proportional to the temperature of 
the phononic bath in the leads, $\theta^2 \sim T_L$ (in agreement  
with Einstein's relation), while it reduces to 1 for the $w(A)$ that describes 
the quantum fluctuations of the oscillator in the ground state.

The denominator of the ratio that defines $\theta^{-2}$ describes the diffusion 
in the energy space of the oscillator induced by the stochastic electronic 
tunnelling processes, whereas the numenator represents the effective damping 
generated by them (see Eq. (\ref{def:theta})). 

In order to understand the physical origin of the stationary state of the 
oscillator determined by the temperature-activated electron tunnelling, let us 
consider the limit $Q\rightarrow \infty$, that is the situation in which the 
nanotube is decoupled from the equilibrium environment. In this limit the 
quasi-distribution function is determined only by the rates of the inelastic 
tunnelling transitions induced by the polaronic and tunnelling electromechanical 
couplings, which are characterized by the parameters $\varepsilon_p$ and 
$\varepsilon_t$, respectively. The stationary state in this limit is characterized 
by a width given by
\begin{equation} \label{eq:theta-Q_inf}
\theta_{Q\rightarrow \infty}^2 = \sqrt{\frac{1}{2}(1+\Gamma_\Sigma^2 )}
\left(\eta + \frac{1}{\eta}\right),
\end{equation} 
\noindent where 
$\eta = \sqrt{2\varepsilon_p^2 / \varepsilon_t^2(1+\Gamma_\Sigma^2)}$. 
It follows from Eq. (\ref{eq:theta-Q_inf}) that for both strong and weak
polaronic coupling $\theta$ is larger than 1, which corresponds to a state
that is far from the quantum ground state. Nevertheless, in a suitable 
range of values of $\varepsilon_p$, $\theta^2$ is significantly smaller 
than $\coth(\hbar \omega_0 /2k_BT_L)$, which means that the stationary 
state can be interpreted as a thermal state characterized by an effective 
temperature smaller than $T_L$. Correspondingly, the root-mean-square 
fluctuations of the nanotube center-of-mass position are smaller than 
the value determined in the thermal equilibrium case, therefore the mechanical 
subsystem is effectively \textit{cooled}. The minimum value that $\theta^2$ 
can reach as a result of the interaction of the nanotube with the 
nonequilibrium electronic environment is given by 
$\theta_{min}^2 = \sqrt{2(1+\Gamma_\Sigma^2)}>1$.

From Eq. (\ref{eq:theta-Q_inf}), one can see that the effective cooling of
the mechanical vibrations induced by the nonequilibrium environment 
requires the presence of both the mechanisms of electromechanical coupling 
in order to exist. In the case in which only one mechanism is active, it 
follows from Eq. (\ref{def:theta}) that it can only generate diffusion in 
energy space, which results in a broadening of the quasi-distribution 
function. We stress that this behavior is characteristic of the 
nonequilibrium situation considered here, since in the case of coupling
with an equilibrium environment, both damping and diffusion are present, 
independently of the type of interaction.

\begin{figure}
\center
\includegraphics[width=0.6\textwidth]{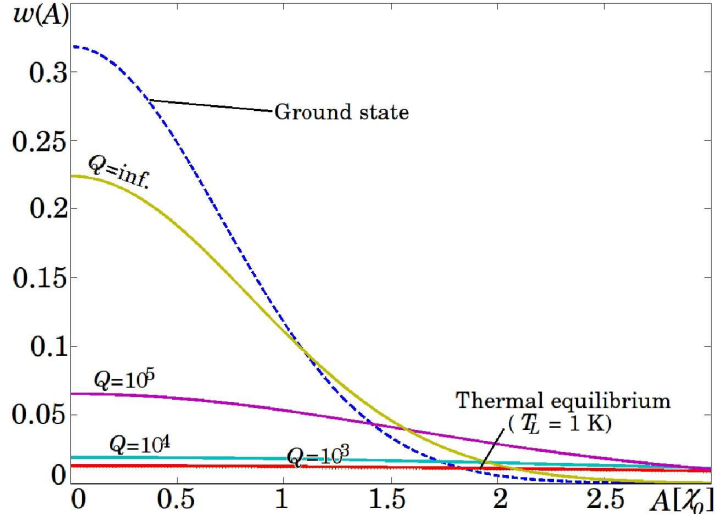}
\caption{Comparison between the Wigner functions corresponding to the ground 
state of the oscillator, the thermal equilibrium state at temperature $T_L$ and
the stationary state induced by the nonequilibrium electronic environment for 
different quality factors. Values of the relevant parameters: 
$\omega_0 = 10^9$ Hz, $\Gamma_S = 5 \cdot 10^6$ Hz, $\Gamma_L = 10^8$ Hz,
$\varepsilon_t = 0.19$, $\varepsilon_p = 0.14$.}
\label{fig:comparison_W}
\end{figure}

\begin{figure}
\center
\includegraphics[width=0.6\textwidth]{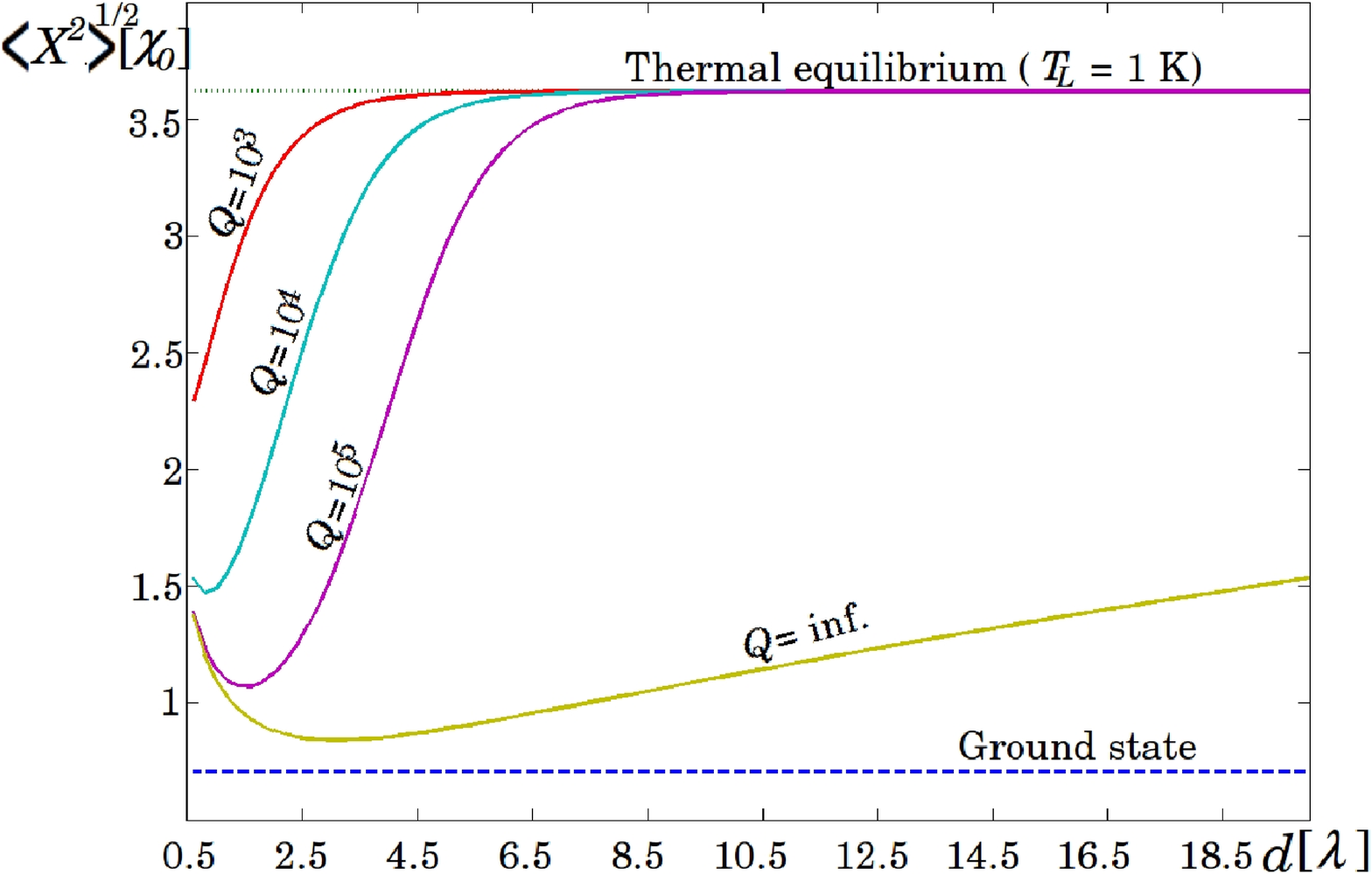}
\caption{Comparison between the fluctuations of the root-mean-square
fluctuations of the center-of-mass nanotube position
$\sqrt{\langle X^2 \rangle}$ calculated with the
Wigner functions corresponding to the ground state of the
oscillator, the thermal equilibrium state at temperature
$T_L$ and the stationary state induced by
the nonequilibrium electronic environment as a function of
the equilibrium distance between the STM tip and the nanotube,
$d$, for different quality factors. Values of the relevant parameters:
$\omega = 10^9$ Hz, $\Gamma_S = 10^8$ Hz, $\Gamma_L = 10^8$ Hz,
$\lambda = 10^{-10}$ m, $\varepsilon_t = 0.19$.}
\label{fig:comparison_rmsX}
\end{figure}

The stationary quasi-distribution function generated by the temperature 
drop from the STM tip to the leads is plotted with respect to the 
amplitude $A$ (expressed in units of $\chi_0$) in 
Fig. \ref{fig:comparison_W}, for different values of the quality factor.  
\noindent Furthermore, we can compare the size of the root-mean-square 
fluctuations of the nanotube center-of-mass position in the ground state,
the thermal equilibrium state and the electronically-induced stationary state 
as a function of the equilibrium distance between the nanotube and the 
STM tip, $d$, for different quality factors, as shown in 
Fig. \ref{fig:comparison_rmsX}.
\noindent The curves shown in Figs. (\ref{fig:comparison_W}) and
(\ref{fig:comparison_rmsX}) indicates that the interaction between 
the electron tunnelling current and the oscillator can be interpreted 
as an effective cooling of the mechanical degrees of freedom. 

In order to detect experimentally the cooling effect predicted above, 
the most direct approach consists in the measurement of the root-mean-square 
fluctuations of the nanotube center-of-mass position. Regarding this point, 
it is has been argued since a long time ago that the STM (combined with a 
current amplifier) can provide the basic building block for a quantum-limited 
position displacement sensor. The tunnelling current that can be measured at 
the output of such a device contains information about the displacement of the 
mechanical system under investigation but, at the same time, perturbs it with 
a very small back-action force, being this mainly due to the random momentum 
transfer associated with the tunnelling electrons \cite{Bocko1988}. 

In conclusion, we have studied the coupled dynamics of the mechanical and
electronic degrees of freedom of a suspended-nanowire-based NEMS wherein 
the movable element is in tunnelling contact with the tip of an STM and two 
supporting metallic leads. Our analysis shows that, in the regime in which 
the electron transport is activated only by the temperature difference 
between different parts of the device, an effective cooling of the mechanical 
degrees of freedom can be achieved. This result depends crucially on the 
interplay between tunnelling and electrostatic coupling that characterizes 
the system considered here. The interaction of the suspended nanowire with 
the nonequilibrium environment provided by the tunnelling current reduces 
the amplitude of the root-mean-square fluctuations of the center-of-mass 
position of the nanowire. This effect could be in principle detected 
experimentally thanks to the almost-quantum limited sensitivity of the STM 
as a displacement sensor.

%The reduced back-action makes it possible for the STM-based displacement 
%sensor to achieve a sensitivity that is several orders of magnitude larger 
%than that one of conventional (e.g. capacitive) electromechanical transducers.
\section*{Acknowledgments}
Partial support from the Swedish VR and SSF, the EC 
project QNEMS (FP7-ICT-233952), the Faculty of Science 
at the University of Gothenburg through its Nanoparticle 
Research Platform, and the Korean WCU program funded 
by MEST/NFR (R31-2008-000-10057-0) is gratefully acknowledged.

%-----------------------------------------
%\bibliography{/home/santandr/Papers/MyDataBase}

\end{document}